\pdfoutput=1

\documentclass[11pt]{article}

\usepackage[final]{emnlp2021}

\usepackage{times}
\usepackage{latexsym}

\usepackage{times}
\usepackage{latexsym}
\usepackage{verbatim}
\usepackage{booktabs}
\usepackage{float}

\usepackage[T1]{fontenc}

\usepackage[utf8]{inputenc}

\usepackage{microtype}
\usepackage{subcaption}
\usepackage{mathtools}
\usepackage{multirow}
\usepackage{amsmath,amsfonts,bm}
\usepackage{amssymb}
\usepackage{bbm}
\usepackage{algorithm}
\usepackage{algpseudocode}
\usepackage{algcompatible}
\usepackage{todonotes}
\usepackage{xspace}

\usepackage{hyperref}
\usepackage{url}
\usepackage{todonotes}
\usepackage{array}

\definecolor{forestgreen}{rgb}{0.13, 0.55, 0.13}

\title{Flood Event Extraction from News Media\\ to Support Satellite-Based Flood Insurance}

\author{Tejit Pabari\textsuperscript{1}, Beth Tellman\textsuperscript{2}, Giannis Karamanolakis\textsuperscript{1}, Mitchell Thomas\textsuperscript{3}, Max Mauerman\textsuperscript{2}, \\ Eugene Wu\textsuperscript{1}, Upmanu Lall\textsuperscript{4}, Marco Tedesco\textsuperscript{3}, Michael S Steckler\textsuperscript{3}, Paolo Colosio\textsuperscript{5}, \\ Daniel E Osgood\textsuperscript{2}, Melody Braun\textsuperscript{2}, Jens de Bruijn\textsuperscript{6}, Shammun Islam \\ \\
\textsuperscript{1}Computer Science Department, Columbia University in the City of New York\\
\textsuperscript{2}International Research Institute for Climate and Society, Columbia University in the City of New York\\
\textsuperscript{3}Lamont-Doherty Earth Observatory, Columbia University in the City of New York\\
\textsuperscript{4}Department of Earth and Environmental Engineering, Columbia University in the City of New York\\
\textsuperscript{5}University of Brescia, Italy, $\:$ 
\textsuperscript{6}Vrije Universiteit Amsterdam, Netherlands}

\begin{document}
\maketitle
\begin{abstract}
Floods cause large losses to property, life, and livelihoods across the world every year, hindering sustainable development. 
Safety nets to help absorb financial shocks in disasters, such as insurance, are often unavailable in regions of the world most vulnerable to floods, like Bangladesh. 
Index-based insurance has emerged as an affordable solution, which considers weather data or information from satellites to create a ``flood index'' that should correlate with the damage insured.
However, existing flood event databases are often incomplete, and satellite sensors are not reliable under extreme weather conditions (e.g., because of clouds), which limits the spatial and temporal resolution of current approaches for index-based insurance.

In this work, we explore a novel approach for supporting satellite-based flood index insurance by extracting high-resolution spatio-temporal information from \emph{news media}. 
First, we publish a dataset consisting of 40,000 news articles covering flood events in Bangladesh by 10 prominent news sources, and inundated area estimates for each division in Bangladesh collected from a satellite radar sensor.
Second, we show that keyword-based models are not adequate for this novel application, while context-based classifiers cover complex and implicit flood related patterns.
Third, we show that time series extracted from news media have substantial correlation (Spearman's $\rho$=0.70) with satellite estimates of inundated area. 
Our work demonstrates that news media is a promising source for improving the temporal resolution and expanding the spatial coverage of the available flood damage data.
\end{abstract}

\begin{figure}[t]
    \centering
    \includegraphics[width=\columnwidth]{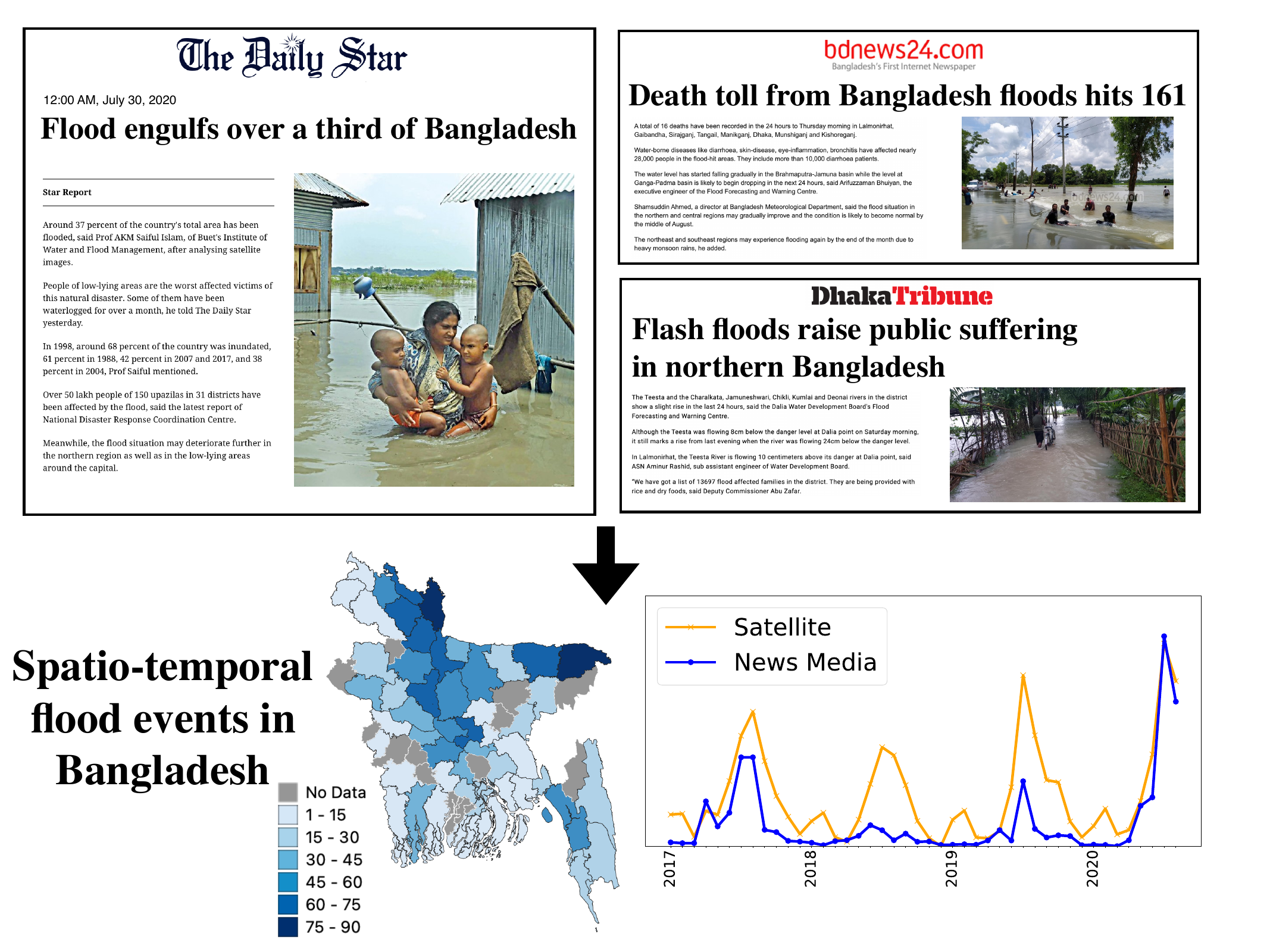}
    \caption{News media provide fine-grained spatio-temporal information about flood events, which can be utilized to facilitate policies for regions of the world that are most vulnerable to floods, such as Bangladesh.}
    \label{fig:intro-flood-articles}
\end{figure}

\section{Introduction}
Flood events affect millions of people and cause billions in damages across the world every year \citep{guha2016dat}. 
Traditional flood insurance for financial protection from these damages involves contracts where adjusters visit individual homes to assess damages and determine corresponding payouts.
However, traditional insurance is often unavailable in regions of the world most vulnerable to floods, such as Bangladesh.
Flood index insurance is an affordable and scalable alternative whereby a data source, often from weather stations or satellites, generates a ``flood index'' (e.g., inundated area) that should correlate with the damage insured~\citep{surminski_how_2016}. 

Public satellite sensors are currently being piloted for flood index insurance in Southeast Asia \citep{matheswaran_flood_2019}. While satellites can map large and slow moving floods, inadequate revisit times\footnote{Revisit time is the time period elapsed between consecutive images of the same point on earth captured by a satellite.} or cloud cover can make mapping maximum flood extent elusive, especially in flash flood events \citep{hawker_comparing_2020, alfieri_global_2018}. Designing a good satellite-based index requires quantifying this uncertainty in flood estimation with reliable spatio-temporal damage data \cite{benami_uniting_2021}. Unfortunately, flood event databases such as EM-DAT \citep{guha2016dat} and Munich Re's NatCatSERVICE \citep{Kron2012} are incomplete, do not contain data at a high spatial or temporal resolution, and often lack small, local or frequent but damaging flood events (especially flash floods \citep{gaillard_alternatives_2010}). 

In this work, we explore a novel approach of using \emph{news media} to extract high-resolution spatio-temporal information about flood events to support flood index insurance (see Figure~\ref{fig:intro-flood-articles}).
In contrast to previous approaches based on social media \citep{de2019global, arthur_social_2018}, we explore news media which cover a broader time range and provide a more detailed account of property and agricultural damage caused by floods.
Our work focuses on Bangladesh, one of the countries most vulnerable to floods.
We collect and annotate news articles from ten prominent news sources, evaluate several techniques for the extraction of flood events, and use the extracted news media information to create time-series for comparison to international disaster databases and satellite data.
Overall, we make the following contributions: 
\begin{itemize}
    \item We propose a novel application of news media for spatio-temporal flood event extraction. To support this application, we collect a dataset of news articles with flood event annotations, and satellite inundation data.\footnote{We plan to publish our dataset and Python code that are both provided in the supplementary material.} 
    \item We perform an extensive evaluation of several flood event classification techniques and show that keyword-based approaches are not adequate for this novel application, while context-based classifiers capture more complex expressions and achieve higher performance.
    \item We use the extracted information to construct time series and show a statistically significant correlation (Spearman's $\rho$=0.70, $p$<1.0E-6) with time series from satellite sensors detecting inundated area, outperforming correlations between Twitter and satellite data (Spearman's $\rho$=0.55, $p$<0.001). 
\end{itemize}
Crucially, our work demonstrates the potential of news media analysis for the accurate extraction of flood events on a finer-grained spatial and temporal resolution compared to official disaster databases and at higher quality than Twitter event extraction.

\section{Related Work}
Our work is related to approaches for disaster management, including disaster type detection~\citep{palen2010vision,sakaki2010earthquake,stowe2016identifying,li2017data} and disaster summarization~\citep{li2006extractive,wang2010document,guo2013updating,kedzie2015predicting}. 
There is substantial work on the analysis of social media for disaster-related discussions~\citep{imran2015processing,imran2016twitter,stowe2016identifying,nazer2017intelligent,nguyen2017robust,hasan2019real,hiware2020narmada,chowdhury2020cross,wiegmann2021opportunities}.
Such discussions might sometimes start even before official warnings are available~\citep{ijgi4042246}, and therefore, social media is promising for the rapid response to emerging events such as earthquakes~\citep{sakaki2010earthquake,alam2018domain,desai2020detecting}, hurricanes~\citep{stowe2018improving} and floods \citep{de2019global,moore2020using}. 
However, social media are characterised by informal short text that is not fact-checked by journalists, and therefore approaches for flood event extraction using Twitter~\citep{de2019global,moore2020using} may not be appropriate for supporting index insurance for agricultural applications in rural areas \citep{duggan2013demographics}.

Since our application requires more reliable information, especially in rural areas, we focus on news articles published by prominent news media sources.
Previous work uses news media to detect various events~\citep{piskorski2020new}, such as socio-political events~\citep{hurriyetouglu2020automated}, there is limited work on the detection of environmental disasters~\citep{nugent2017comparison}. 
Here we explore the potential of news media for the extraction of spatio-temporal flood events and in addition to document classification, we compare news-extracted time series to satellite data.

\section{Dataset Creation}
In this section, we summarize our dataset consisting of news articles and spatio-temporal flood damage data from a satellite sensor.
See the Appendix for detailed descriptions of the collection process and dataset statistics.

\subsection{Article Collection and Annotation}
We focus on English\footnote{Considering non-English articles would introduce several extra challenges that we discuss in Section~\ref{s:discussion}.} news articles about flood events in Bangladesh that were published in the websites of ten different news sources: nine prominent national Bangladesh news media (``The Daily Star,'' ``BD News,'' ``Daily Observer,'' ``Daily Sun,'' ``Dhaka Tribune,'' ``New Age,'' ``Prothomalo,'' ``The Independent,'' ``The New Nation''), and the international ``New York Times.'' 
In total, we have collected 39,777 articles that have been published on a date between 2000 and 2020 and match our search queries with flood-related keywords (see the Appendix for details).

We employed two flood domain experts to classify news articles associated with historical flood events in Bangladesh (and identify articles that did not meet this criteria) for a subset of our dataset. 
In total, 1380 articles have been annotated with binary (400 ``flood'', 980 ``not flood'') labels on whether the article mentions a \emph{real} flood event (and not hypothetical flood events or mitigation policies). 

\subsection{Satellite Data Collection}
\label{s:satellite-data-collection}
To evaluate the informativeness of news media as a source for flood event detection we augment our dataset with spatio-temporal data estimated by an algorithm using the Sentinel-1 satellite sensor via the Google Earth Engine platform~\citep{devries_rapid_2020}.
Data are stored as weekly inundated area estimates for each division in Bangladesh from 2017 to 2020.

\section{Flood Event Classification}
\label{s:flood-classification-models}
We evaluate various techniques for binary flood event classification (i.e., predicting whether an article discusses a flood event):
 a simple heuristic based on flood-related keywords\footnote{We report the best keyword-based classifier: if at least one of \textit{flood(s)}, \textit{inundat(ion)(ed)}, or \textit{cyclone(s)} appears in the document, the prediction is ``flood'', otherwise it is ``not flood.'' See the Appendix for more approaches.},  bag-of-words classifiers (Logistic Regression, Support Vector Machine or SVM, and Random Forest), and a classifier based on pre-trained BERT representations~\citep{devlin2019bert}.
 For details on the model configuration, experimental procedure, and evaluation results, see the Appendix. 

Table~\ref{tab:flood-classification-results} reports the classification performance results computed on a held-out test set of 880 annotated articles.
Relying on simple keyword-based heuristics leads to a substantially lower F1 score than machine learning approaches: many articles contain flood-related keywords but do not discuss real flood events (e.g., ``\textit{The season of flood, cyclone and dengue is upcoming}''), while there is a substantial number of articles that do not contain keywords but refer to flood events through more complex descriptions (e.g., ``\textit{... many other parts of the capital went under the knee-to-waist-deep water, causing immense sufferings to the city dwellers}'').
Our results highlight the need for techniques that effectively leverage the rich context in news articles to detect real flood events.
The highest F1 score is achieved by the BERT-based classifier: leveraging pre-trained BERT embeddings for the articles is more effective than relying on simple bag-of-words representations (Logistic Reg., SVM, Random Forest). Only BERT was capable of identifying flood events described in context with no obvious keywords (e.g ``\textit{Water has seeped into households...}").

\begin{table}[t]
    \centering
\begin{tabular}{lcc}
        \toprule
        \textbf{Method} & \textbf{Accuracy} &\textbf{F1}   \\
        \midrule
       Keywords &73.8 & 77.7 \\
    Logistic Regression &90.5 & 90.2\\ 
    SVM  & 91.4 & 91.1\\ 
    Random Forest & 91.7 & 91.8\\ 
    BERT & \textbf{93.8} & \textbf{93.6}\\ 
     \bottomrule 
    \end{tabular}
\caption{Results for binary flood event classification. }
    \label{tab:flood-classification-results}
\end{table}

\begin{table*}[t]
    \centering
    \resizebox{2\columnwidth}{!}{
    \begin{tabular}{l|l|l|llllllll}
        \toprule 
        & \textbf{EM-DAT} & \multicolumn{9}{c}{\textbf{Satellite Data for Bangladesh and its 8 Divisions}} \\
       & Bangladesh & Bangladesh & Sylhet &  Rajshahi &  Dhaka &  Barisal &  Chittagong &  Khulna &  Rangpur & Mymensingh  \\
        \midrule
        \midrule
        Twitter  & 0.27* & 0.55*** &0.69*** &   \textbf{0.46**} &  0.47** &  0.24 &     0.51*** &  0.29* &  0.34* & 0.34*\\
        News Media & \textbf{0.44***} & \textbf{0.70***} &\textbf{0.76***} &   0.41** &  \textbf{0.65***} &  \textbf{0.59***} &     \textbf{0.67***} &  \textbf{0.54***} &  \textbf{0.34**} & \textbf{0.59***}\\
        \bottomrule
    \end{tabular}}
    \caption{Comparison between time series constructed by Twitter vs. news media in cross-correlation (Spearman's $\rho$) with the EM-DAT database (n=5) and satellite data (n=44). Results marked as statistically significant at the p<0.1*, p<0.05**, and p<0.01*** levels. We also report Pearson correlation results in the Appendix.}
    \label{tab:all-crosscorr-results-spearman}
\end{table*}
\section{Spatio-Temporal Flood Event Analysis}
We evaluated news media as a source for flood event analysis at a fine-grained temporal and spatial resolution, comparing to Twitter~\citep{de2018taggs,de2019global}. Spatial coverage of flood events extracted from news was extensive; 52 (\%81) of districts had > 1 flood event detected (Fig. 1).

We create time-series from news media based on the assumption that a flood event is severe if there is high redundancy in articles covering this event.
First, we apply the best classifier (BERT) from Section~\ref{s:flood-classification-models} across all 40,000 articles in our dataset to get ``flood'' vs. ``not flood'' predictions.
Additionally, for each article, we extract the date and the location of the flood event (see the Appendix for implementation details).
Then, we compute the resulting time series value for a given date range (e.g., week) and location (e.g., whole Bangladesh, or division of Bangladesh) as the number of articles classified as ``flood'' divided by the estimated\footnote{We estimate this number by scraping news sources for all articles within the given date range and location that contain the keyword ``the''.} number of total articles published.

We compared Twitter and news media data to satellite inundation estimates from Section~\ref{s:satellite-data-collection} and the global disaster event database EM-DAT~\citep{guha2016dat} in Table~\ref{tab:all-crosscorr-results-spearman}.
EM-DAT represents an independent set of large flood events but excludes smaller events.\footnote{We used the estimated number of people affected by floods in Bangladesh in EM-DAT (n=5) as a proxy for severity.} Our results reveal the improved information content of news media for flood event extraction compared to Twitter. 
First, flood events from news media are more highly correlated with EM-DAT data. Second, news media show higher correlation to satellite data.  Twitter shows lower correlation than news media in most areas because it is not fact-checked and is biased towards urban areas. Therefore, Twitter may be less appropriate for understanding agricultural index insurance, which needs high accuracy and coverage in rural areas.

\begin{figure}
    \centering
        \begin{subfigure}[t]{\columnwidth}
        \includegraphics[width=\columnwidth]{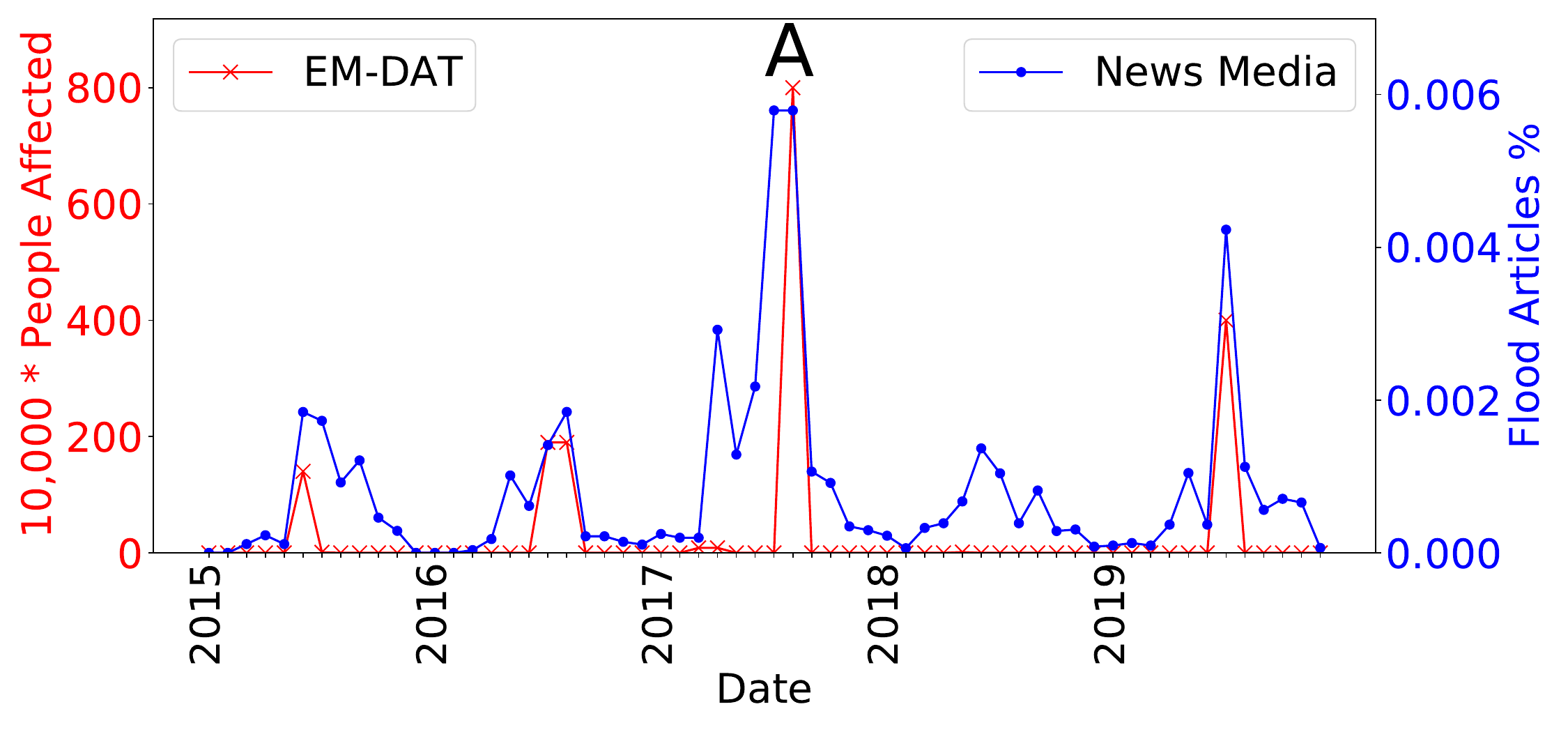}
        \caption{News media to EM-DAT}
        \label{fig:flood-time-series-emdat}
        \end{subfigure}
        \begin{subfigure}[t]{\columnwidth}
        \includegraphics[width=\columnwidth]{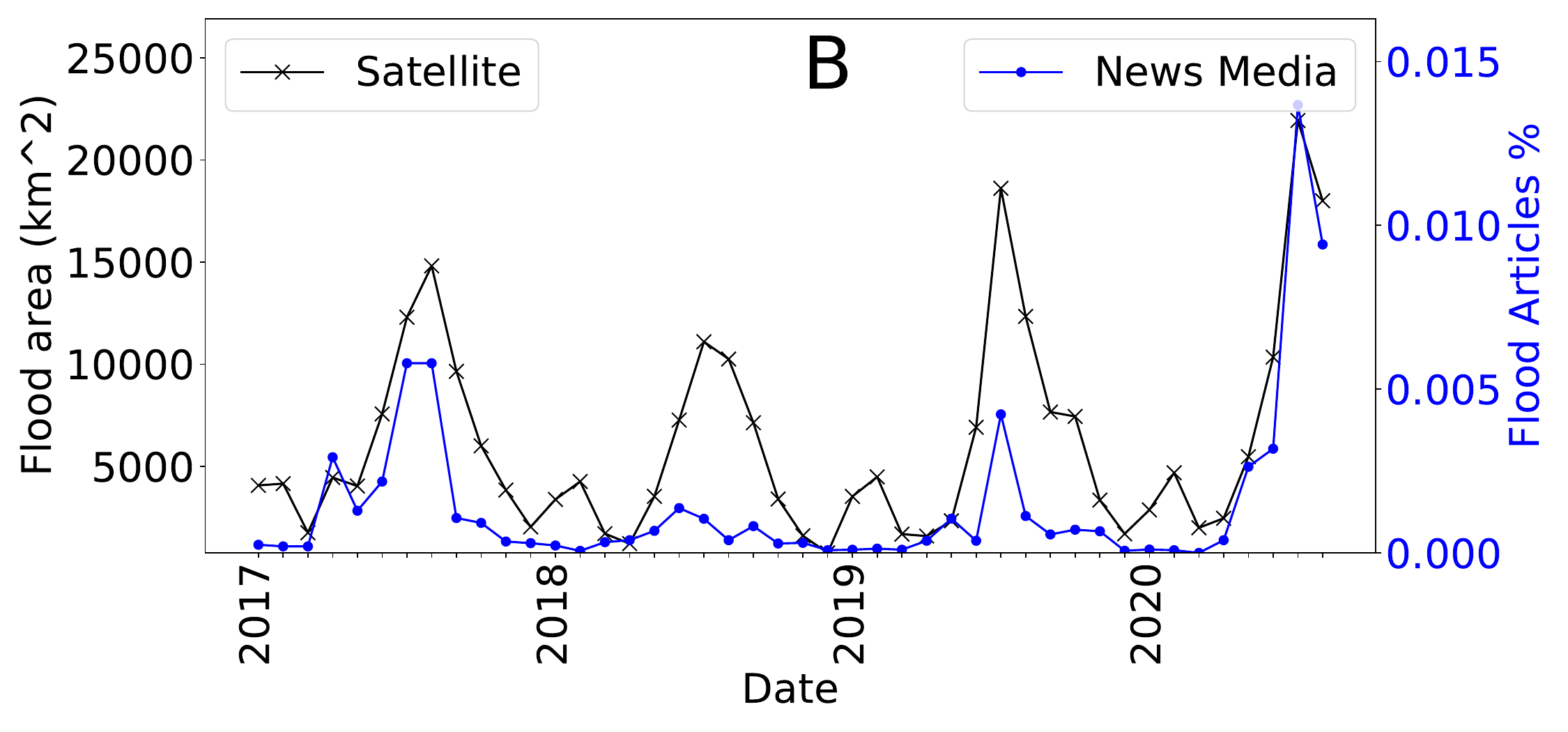}
        \label{fig:flood-time-series-satellite}
        \hypertarget{fig:flood-time-series-satellite}{}
        \end{subfigure}
        \vspace{-1\baselineskip}
    \caption{Flood event time series comparison: (a) news media to EM-DAT; (b) news media to satellite.}
    \label{fig:flood-time-series}
\end{figure}

Figure~\ref{fig:flood-time-series} shows the news media time series compared to EM-DAT and satellite.
In Figure~\ref{fig:flood-time-series-emdat}, news media time series capture the ``peaks'' of the EM-DAT time series. As expected, the number of news articles increased along with the number of people exposed.
However, news media articles cover smaller events important for assessing flood damages not reported in EM-DAT.
In Figure~\hyperlink{fig:flood-time-series-satellite}{2b}, we observe substantial correlation of flood event information from news media versus satellite estimates.

\section{Discussion and Future Work}
\label{s:discussion}
We presented a novel application of news media to extract flood events on a fine-grained spatial and temporal resolution.
Besides publishing a new dataset with news articles covering flood events in Bangladesh, we demonstrate that deep learning classifiers (BERT with the highest extraction accuracy) lead to news-article time series that have a statistically significant correlation with time series from satellite sensors. 
Our results serve as a proof-of-concept that news media are valuable for understanding index insurance proxies such as satellite data and flood forecasting applications~\citep{coughlan_de_perez_forecast-based_2015}: news media provide information about small events that do not appear in official databases, have higher spatial resolution coverage by mentioning specific regions that official data do not capture, and are less biased and more consistent than social media.

As a limitation of our work, our classifiers can be applied for articles written in English but not articles in the local language (Bengali), which could potentially disclose more detailed information about flood events. 
Training supervised classifiers for Bengali articles would require to employ flood experts who are native Bengali speakers for the annotation of thousands of articles, which is beyond the scope of this paper and our budget. 
One interesting future direction for detecting flood events for any target language is to employ cross-lingual text classification and train target classifiers using our annotated English dataset and unlabeled target language articles~\citep{ruder2019survey}.
Future work also includes strategies to effectively aggregate different news sources, techniques for multi-modal extraction, and identification of flood severity.
\section*{Ethical Considerations}
In this work, we introduce the application of spatio-temporal flood event extraction from news media.
Through the empirical evaluation of several classification techniques on our expert annotated dataset, we show that flood events can be accurately detected from the text in news articles and we constructed time series that significantly correlate to external data sources such as satellites. 
To this end, our dataset and classifiers can be used for applications in disaster management, environmental studies, land use planning, finance, legal, healthcare and other domains where there is interest in the analysis of flood events. One limitation of this dataset is that news media does not cover all of the flood events that occur in Bangladesh, even though it does a better job covering small, local events in rural areas than official data or Twitter. We suspect many small events that devastate livelihoods remain unrepresented in our dataset because media did not cover them, especially in remote locations. The main annotator in this project was financially compensated at academic research assistant rates and co-authored the paper. The second annotator did not require compensation as this fits under their academic research activities, and also co-authored the paper.

While our application demonstrates the potential of news media for flood event extraction, it also suffers from associated societal implications of insurance as a a solution to the climate crisis. The most vulnerable populations usually live in floodplains or in "char" islands in the middle of braided rivers and may not own land. For example, in Bangladesh, even low insurance premiums index insurance could enable (e.g., $1$ per year) could be too high for these populations. Agricultural insurance may place the financial burden of reducing vulnerability on the poorest farmers, even though companies in other countries causes the pollution that increased climate change and flood impacts. Other mechanisms, like Loss and Damage \citep{huq_loss_2013}, would place the financial burden on richer, polluting countries to directly pay excess damage costs to Bangladesh to distribute to farmers. We actively engage in these discussions and critiques by discussing with local Bangaldeshi practitioners (NGOs, academics, and insurance companies) how insurance is not a silver bullet but must be used in tandem with other disaster risk reduction strategies. We share our dataset, code, and results with a Bangladeshi science advisory board that has given input on this project. 

\newpage
\bibliography{custom,ref,Bethsrefs}
\bibliographystyle{acl_natbib}

\end{document}